\documentclass[iop,apj]{emulateapj}
\begin{document}
\newcommand{\beq}{\begin{equation}}
\newcommand{\eeq}{\end{equation}}
\def\eeqno#1{\label{#1}\end{equation}}
\def\div{\vec\nabla\cdot}
\def\grad{\vec\nabla}
\def\curl{\vec\nabla\times}
\def\rar{\rightarrow}
\def\re{R_{e}}
\def\az{a_0}
\def\cmst{~{\rm cm~ s}^{-2}~}
\def\gcmt{~{\rm g~ cm}^{-2}~}
\def\vinf{V_{\infty}}
\def\Halpha{H$_{\alpha}~$}
\def\Sz{\Sigma_0}
\def\s{\sigma}
\def\a{\alpha}
\def\b{\beta}
\def\l{\lambda}
\def\deg{^o}
\def\kpc{{\rm kpc}}
\def\mpc{{\rm Mpc}}
\def\gpc{{\rm Gpc}}
\def\St{\Sigma_t}
\def\vg{{\bf g}}
\def\vh{{\bf h}}
\def\vr{{\bf r}}
\def\vR{{\bf R}}
\def\_#1{_{\scriptscriptstyle #1}}
\newcommand{\mss}{\ensuremath{\mathrm{m}\,\mathrm{s}^{-2}}}
\newcommand{\kms}{\ensuremath{\mathrm{km}\,\mathrm{s}^{-1}}}
\newcommand{\MLsun}{\ensuremath{\mathrm{M}_{\sun}/\mathrm{L}_{\sun}}}
\newcommand{\Lsun}{\ensuremath{\mathrm{L}_{\sun}}}
\newcommand{\Msun}{\ensuremath{\mathrm{M}_{\sun}}}
\newcommand{\Aunits}{\ensuremath{\mathrm{M}_{\sun}\,\mathrm{km}^{-4}\,\mathrm{s}^{4}}}
\newcommand{\surfdens}{\ensuremath{\mathrm{M}_{\sun}\,\mathrm{pc}^{-2}}}
\newcommand{\etal}{et al.}
\newcommand{\LCDM}{$\Lambda$CDM}
\newcommand{\ML}{\ensuremath{\Upsilon_*}}
\newcommand{\mmmm}{Paper I}

\title{Andromeda Dwarfs in Light of MOND. II. Testing Prior Predictions}
\author{Stacy  McGaugh}
\affil{Department of Astronomy, Case Western Reserve University, Cleveland, OH 44106, USA}

\and

\author{Mordehai Milgrom}
\affil{Department of Particle Physics and Astrophysics, Weizmann Institute of Science, Rehovot 76100, Israel}

\begin{abstract}
We employ recently published measurements of the velocity dispersions in the newly discovered dwarf satellite galaxies of Andromeda
to test our previously published predictions of this quantity.  The data are in good agreement with our specific predictions for each dwarf made
\textit{a priori} with MOND, with reasonable stellar mass-to-light ratios, and no dark matter, while Newtonian dynamics point to quite large mass 
discrepancies in these systems.  MOND distinguishes between regimes where the internal field of the dwarf, or the external field of the host, dominates.  
The data appear to recognize this distinction, which is a unique feature of MOND not explicable in \LCDM.
\end{abstract}

\keywords{dark matter --- galaxies: kinematics and dynamics --- Local Group}

\section{Introduction}
\label{intro}

The giant and satellite galaxies of the Local Group offer various ways of testing the predictions of \LCDM\ and MOND.
The number statistics of satellites around the Milky Way and Andromeda is one test of the \LCDM\ paradigm, leading to 
the well known missing satellite problem \citep[e.g.,][]{Moore1999}.
More recently, the phase-space distributions of the satellites around their host galaxy have been shown to be an additional challenge for \LCDM\ \citep{kroupafalse}.
Rather than resembling the orbit distribution of sub-halos in structure formation simulations, many of the satellites are organized into rather thin, rotating disks,
as shown for the Milky Way by \citet{pawlowski12} and for M31 by \citet{ibata13}.  Whether the dSph satellites of the Local Group are
primordial dwarfs residing in sub-halos or tidally formed remnants is another possible test to distinguish \LCDM\ and MOND \citep[e.g.,][]{gentiletidal}.
For example, \citet{LGtimingMOND} have made a detailed account of the timing argument in MOND, showing that a near collision between the Milky Way
and Andromeda $\sim$10 Gyr ago might lead to the formation of the thick disk and the observed properties of the disks of satellites.
On the other hand, we have perhaps been wrong to assume that dSph galaxies are all primordial residents of Galactic sub-halos, as some might arrive through
filamentary accretion \cite{stewart13,wang13}.  This might explain the observed phase space distribution in \LCDM\ while also easing the challenge of 
suppressing dwarf formation in sub-halos without squelching it entirely \citep{Somerville2002}.
\citet{ShayaTully} find that the observed planarity arises fairly naturally in reconstructions of the orbits of Local Group galaxies that fit the available observations, 
though it is not obvious that the smooth density component that they invoke is consistent with the expectations of structure formation simulations.

The amplitude of the mass discrepancy in each galaxy provides another test.
In a recent analysis \citep[hereafter \mmmm]{mm13}, we considered the internal dynamics of the dSph satellites of Andromeda.
We employed MOND \citep{milg83} to calculate the velocity dispersion of stars in each of the dwarf spheroidal satellites of
Andromeda, many of which have only recently been discovered.

MOND posits that the mass discrepancies observed in galactic systems are not the result of dark matter,
but rather of modified dynamics at low accelerations $a \lesssim \az \approx 10^{-10}\;\mathrm{m}\,\mathrm{s}^{-2}$.
In MOND the dynamics in the limit $a\ll\az$ are required to be invariant under space-time scaling,
$(t,\mathbf{r}) \rightarrow \lambda(t,\mathbf{r})$ \citep{milgrom09}.
The MOND calculations of the expected velocity dispersions in \mmmm\ are full fledged predictions, as they use inflexible procedures,
with no leeway for reckoning with known values of the dispersions.

When we wrote \mmmm, velocity dispersion data were available for seventeen of the dSph satellites of M31.
The predicted velocity dispersions were consistent with the majority of the data within its stated uncertainty.
In a number of cases (most notably And II, but also And I, III, VII, and XIV), the predictions were in good agreement with multiple
independent observations.  In some cases (most notably And V and And IX), the prediction agreed better with one observation than another.
These cases fall in the category of inflexible expectations: the known velocity dispersions play no role in the prediction.

\begin{deluxetable*}{lccccccl}
\tablewidth{0pt}
\tablecaption{Predicted and Observed Velocity Dispersions}
\tablehead{
\colhead{Dwarf} & \colhead{$L_V$} & \colhead{$r_{1/2}$} & \colhead{$\sigma_{iso}$} & \colhead{$\sigma_{efe}$}
 & \colhead{$\sigma_{obs,1}$} & \colhead{$\sigma_{obs,2}$} & \colhead{Ref.} \\
 & \colhead{$10^5\;L_{\sun}$} & pc & \multicolumn{2}{c}{\kms} & \multicolumn{2}{c}{\kms\ (\#)}
}
\startdata
And XVII   & \phn2.6\phn & \phn381 & $4.5^{+0.9}_{-0.7}\pm0.4$ & $2.5^{+1.0}_{-0.7}\pm0.5$ & $2.9^{+2.2}_{-1.9}$ (8) & \dots & 1,2 \\
And XVIII\tablenotemark{a} & \phn6.3\phn     & \phn417 & $5.6_{-0.9}^{+1.1}\pm2.6$  & \dots	& $\le 2.7$ (4) & $9.7\pm2.3$ (22) & 1,2,3 \\
And XIX    & \phn4.1\phn & 2244 & $5.0^{+1.0}_{-0.8}\pm0.7$ & $2.6^{+1.1}_{-0.8}\pm0.8$ & $4.7^{+1.6}_{-1.4}$ (26) & \dots & 1,2 \\
And XX 	& \phn0.28  & \phn165  &  $2.6^{+0.5}_{-0.4}\pm0.5$  & $2.1^{+0.9}_{-0.6}\pm0.9$ & $7.1^{+3.9}_{-2.5}$ (4) & \dots & 1,2 \\
And XXI\tablenotemark{a}   & \phn4.6\phn     & 1023    	 & $5.2_{-0.8}^{+1.0}\pm0.8$  & $3.7^{+1.5}_{-1.1}\pm1.3$ & $4.5^{+1.2}_{-1.0}$ (32) & $7.2\pm5.5$ (6) & 1,2,3 \\
And XXII\tablenotemark{a}  & \phn0.35        & \phn340 	& $2.7_{-0.4}^{+0.5}\pm0.5$  & $2.3^{+1.0}_{-0.7}\pm1.0$ & $2.8^{+1.9}_{-1.4}$ (10) & $3.5_{-2.5}^{+4.2}$ (7) & 1,2,3 \\
And XXIII  & 10.\phn\phn & 1372 & $6.4^{+1.2}_{-1.0}\pm0.7$ & $4.4^{+1.8}_{-1.3}\pm1.0$ & $7.1\pm1.0$ (42) & \dots & 1,2 \\
And XXIV   & \phn0.94 & \phn489 & $3.5^{+0.7}_{-0.6}\pm0.4$ & $2.8^{+1.2}_{-0.8}\pm0.7$ & $\le 7.3$ (12) & \dots & 1,2 \\
And XXV    & \phn6.5\phn & \phn945 & $5.7^{+1.1}_{-0.9}\pm0.7$ & $3.5^{+1.5}_{-1.0}\pm0.8$ & $3.0^{+1.2}_{1.1}$ (26) & \dots & 1,2 \\
And XXVI   & \phn0.59 & \phn296 & $3.1^{+0.6}_{-0.5}\pm0.4$ & $2.0^{+0.8}_{-0.6}\pm0.5$ & $8.6^{+2.8}_{-2.2}$ (6) & \dots & 1,2 \\
And XXVII\tablenotemark{b} & \phn1.2\phn & \phn579 & $3.7^{+0.7}_{-0.6}\pm0.4$ & $1.8^{+0.7}_{-0.5}\pm0.4$ & 14.8 (8) & \dots & 1,2 \\
And XXVIII & \phn2.1\phn & \phn284 & $4.3^{+0.8}_{-0.7}\pm0.9$ & \dots & $6.6^{+2.9}_{-2.1}$ (17) & $4.9\pm1.6$ (18) & 1,2,4 \\
And XXIX   & \phn1.8\phn & \phn481 & $4.1^{+0.8}_{-0.7}\pm0.3$ & $3.8^{+1.6}_{-1.1}\pm0.6$ & \dots & $5.7\pm1.2$ (24) & 1,4
\enddata
\tablerefs{Luminosities and half light radii are from 1.~\citet{McConnachie2012}.
We assume that the 3D $r_{1/2} = 1.33 R_e$ \citep{boom}.
Measured velocity dispersions are adopted from
2.~\citet[always given as $\sigma_{obs,1}$]{collins13},
3.~\citet{tollerud2012}, and
4.~\citet{tollerud2013}.
The number of stars contributing to the measurement is given in parentheses.
}
\tablecomments{The MOND velocity dispersions predicted \textit{a priori} in \mmmm\ are reproduced here
for comparison to the data.  The first column of uncertainties corresponds to the range of assumed $\ML = 2^{+2}_{-1}\;\MLsun$.
This is a convenient depiction but not an uncertainty in the usual sense, as the mean and range of \ML\ is unknown.
The second column is the conventional uncertainty propagating the stated observational errors around an assumed $\ML = 2\;\MLsun$.}
\tablenotetext{a}{Initial velocity dispersion measurements for And XVIII, XXI, and XXII by \citet{tollerud2012} predate \mmmm,
though only that of And XVIII involved more than seven stars.  The predicted velocity dispersions for all the remaining dwarfs in this
table are completely \textit{a priori}.}
\tablenotetext{b}{\citet{collins13} find that And XXVII is not in equilibrium, so the measured velocity dispersion
does not provide a test of the equilibrium prediction.}
\label{origpredtable}
\end{deluxetable*}

\mmmm\ also predicted the velocity dispersions for another ten dwarfs that did not yet have velocity dispersion measurements.
Since that time, data for these dwarfs have appeared in the literature, providing the opportunity to check our predictions.
We compare our predictions to the new observations in \S \ref{results}.
In \S \ref{newdwarfs} we predict the velocity dispersions of several more new dwarfs.
In \S \ref{discuss} we discuss the results, and consider whether they can be understood in the context of dark matter.
We summarize our conclusions in \S \ref{conclude}.

\section{Comparison of Predictions and Observations}
\label{results}

Velocity dispersion data for the ten dSph galaxies predicted \textit{a priori} in \mmmm\ have been obtained by \citet{tollerud2013} and \citet{collins13}.
\citet{tollerud2013} measure the velocity dispersions of And XXVIII and XXIX based on observations of 18 and 24 stars, respectively.
\citet{collins13} provide comprehensive measurements of very nearly all the known dwarfs of Andromeda.
These monumental observational efforts provide the basis for testing our predictions, which are based on the photometric data compiled by \citet{McConnachie2012}.

\begin{figure*}
\epsscale{1.0}
\plotone{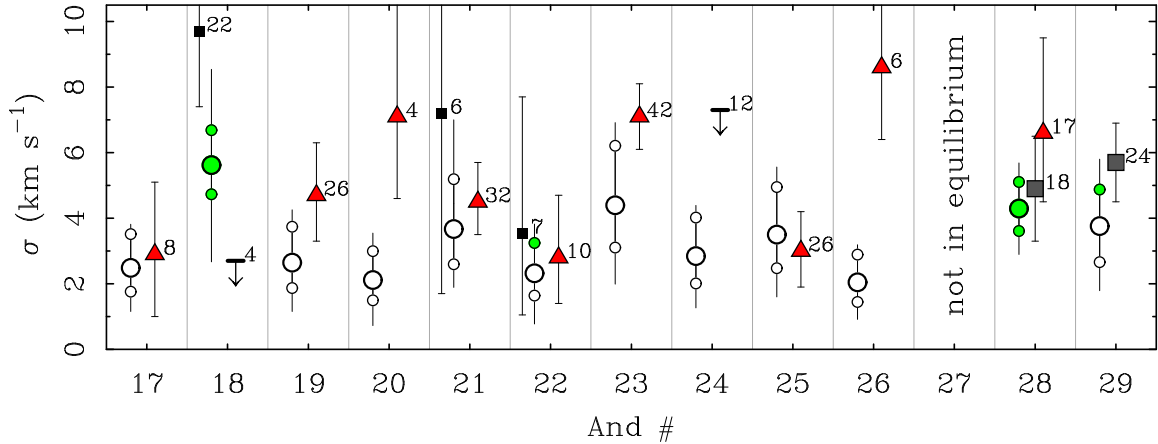}
\caption{The predicted and observed velocity dispersions of recently measured dSph satellites of M31.
Each bin along the x-axis represents the number of the corresponding dwarf for And XVII through And XXIX.
The predictions from \mmmm\ are shown without adjustment as circles for $\ML = 1$ (lower points), 2 (middle points) and $4\;\MLsun$ (upper points).
Circles are filled if the dwarf is in the isolated regime for the assumed mass-to-light ratio and open if it is dominated by the external field of M31.
Error bars propagate all observational uncertainties together with a factor of two uncertainty in the mass-to-light ratio around $\ML = 2\;\MLsun$.
Observed velocity dispersions are taken from \citet[small squares]{tollerud2012}, \citet[large squares]{tollerud2013},
and \citet[triangles or, in the case of upper limits, horizontal bars]{collins13}.  The number of stars used to determine
each velocity dispersion is shown next to the corresponding datum.
Points appear in the chronological order of their appearance in the literature from left to right within each bin.
\citet{collins13} find that And XXVII is not in equilibrium, in contradiction to a necessary assumption of the analysis.}
\label{fig1}
\end{figure*}

The predictions from Table 2 of \mmmm\ are reproduced in Table~\ref{origpredtable} together with the new observations.
In addition to the ten \textit{a priori} cases, we include in Table~\ref{origpredtable} three more dwarfs
(And XVIII, XXI, and XXII) that fall in the numerical sequence and have new or improved data.
We have not adjusted the predictions in any way.  These are compared to the data in Fig. \ref{fig1}.

We calculated the characteristic velocity dispersion of each dwarf galaxy for several assumed values of their stellar mass-to-light ratio:
$\ML = 1$, 2, and $4\;\MLsun$ in the $V$-band. We report this as a plus or minus range in Table~\ref{origpredtable}.  This is a convenient depiction,
but is not an uncertainty in the usual sense.  It simply illustrates the effect of a factor of two variation around a nominal mass-to-light ratio of
$\ML = 2\;\MLsun$, which is a plausible estimate given our present state of knowledge about the stellar populations of
Local Group dwarfs \citep[e.g.,][]{martinmaxL}.

We propagate the observational uncertainties in the photometric properties of the dwarfs into the predicted velocity dispersions.
This observational uncertainty is reported in Table~\ref{origpredtable} separately from the theoretical uncertainty in mass-to-light ratio.
One should of course bear in mind that astronomical uncertainties are rarely well described by conventional random error distributions,
though this is implicit in error propagation.  We are, however, more concerned with the
potential for systematic uncertainties not described by the error bars, such as the trebling of the luminosity of And II (see \mmmm).

\subsection{Isolated and EFE regimes}
\label{isovsefe}

There are two distinct regimes in MOND: that of isolated systems dominated by their internal field and that of systems dominated by an external field
(e.g., that of the host galaxy --- see \mmmm).  In the isolated case, $\s \propto M^{1/4}$, whereas in the case where the external field effect (EFE) 
is important, $\s \propto (M/r g_{ex})^{1/2}$, where $g_{ex} = V_{M31}^2/D_{M31}$ with $D_{M31}$ being the dwarf's distance to M31
and $V_{M31}$ the M31 rotational speed at the position of the dwarf.  The $(M/r)^{1/2}$ dependence is the same as in Newtonian dynamics 
but with an enhanced effective value of Newton's constant that depends on the strength of the external field. 
For each assumed mass-to-light ratio, we compute the velocity dispersion in each of these distinct regimes, and tabulate both
with the subscript $iso$ or $efe$.  In general, the case that applies is the one that returns the lower value for $\s$.  This is reflected in
Fig.~\ref{fig1}, where we illustrate the appropriate choice at each point, and illustrate the distinction with open and filled symbols.

The propagation of observational errors differs between the isolated and EFE case.  Only the luminosity enters the computation of the 
velocity dispersion in the isolated case.  The observational errors contributing to uncertainty in the luminosity are those in the flux measurement
and the distance.  Uncertainties in the magnitudes listed by \citet{McConnachie2012} are typically a few tenths of a magnitude.  These usually
dominate over the formal\footnote{Though extragalactic distance estimates have improved considerably, the distance to individual objects
is still subject to potential systematic errors.  Tabulation errors are also possible:  the literature contains different estimates for the distance to And IX
paired with identical absolute magnitudes.} distance errors.  The computation of the EFE case is less certain, 
as it depends on the size of the dwarf as well as its luminosity, and also on the rotation curve of the host and the distance of the dwarf therefrom (see \mmmm).
We propagate all these errors and report the resulting uncertainty in the velocity dispersion in Table~\ref{origpredtable}.

The predictions velocity dispersions are, by and large, in good agreement with the new observations (Fig.~\ref{fig1}).
Indeed, agreement appears to improve as the number of stars observed in each dwarf increases.
This includes several dwarfs that \citet{tollerud2012} and \citet{collins13} point out as outliers to 
empirical expectations \citep[e.g.,][]{walker,walkerandme,tollerud2011}.
These dwarfs are subject to the EFE, and have predicted velocity dispersions of only a few \kms.
These small velocity dispersions occur naturally in MOND in a way that is not anticipated in the \LCDM\ paradigm.
We return to this important point in \S \ref{paircomp}. 

\subsection{Individual Cases}
\label{indcase}

Here we discuss each individual dwarf in numerical order.
And XVII is the first dwarf that had no published velocity dispersion available for \mmmm.
We therefore start with it, and progress through the last case predicted in \mmmm, And XXIX.
And XVIII, XXI, and XXII had some velocity-dispersion data available in \mmmm, but new data for these cases have appeared
so these are also discussed here.  Dwarfs identified too recently to be included in \mmmm, And XXX, XXXI, and XXXII, are discussed in \S \ref{newdwarfs}.

\subsubsection{And XVII}

The MOND predicted velocity dispersion of And XVII is in good agreement with the newly observed value.
Eight stars, as observed here, is the threshold where velocity dispersions begin to become reliable according to the Monte Carlo
experiments of \citet{collins13}.  Future data with more member stars will make the comparison more stringent.

\subsubsection{And XVIII}

And XVIII is in the isolated regime for all three $\ML$ values. Its predicted velocity dispersion is intermediate between
the measurement of \citet{tollerud2012} and the upper limit obtained by \citet{collins13}.
The latter measurement is based on only four stars, so the apparent conflict between the data is not as serious as it might appear.
Moreover, the large photometric uncertainty in this case spans the difference between the two velocity dispersion measurements.

\subsubsection{And XIX}

The predicted range of velocity dispersions for And XIX overlaps with the subsequently observed range.
There are a good number of stars (26) contributing to this measurement.
It is worth noting that the prediction for the isolated regime is somewhat higher, in even better agreement with the data.
The expected velocity dispersion is clearly in the vicinity of the observed value, but the exact value depends on details like
M31-centric distance, the scale length, and the asymptotic velocity of M31 (see discussion in \mmmm).
It is also worth noting that this is one of three cases (the others being And XXI and And XXV) that \citet{collins13} highlight as large
($R_e > 700$ pc) dwarfs that deviate from empirical mass-luminosity-radius relations \citep[e.g.,][]{tollerud2011}.
These anticipate a higher velocity dispersion than observed, as do the models of \citet{PMN08}. 

\subsubsection{And XX}

The predicted velocity dispersion of And XX is only 2 or 3 \kms, while the observed value is closer to 7 \kms.
However, this value is based on only four stars, so is highly uncertain.

\subsubsection{And XXI}

\citet{tollerud2012} had already reported a velocity dispersion of $\s = 7.2\pm5.5\;\kms$ for And XXI prior to publication of \mmmm.
Their central value is rather larger than our prediction, which is nonetheless within the quoted measurement errors.
This is an example of an inflexible prediction:  we could not arbitrarily match the observed value of $7.2\;\kms$, and out prediction is
only consistent with this measurement because of the large uncertainties.

The velocity dispersion of \citet{tollerud2012} was based on six stars.  More recently, \citet{collins13} report $4.5^{+1.2}_{-1.0}\;\kms$
based on 32 stars.  This is in good agreement with the MOND prediction.  And XXI is another case of a large dwarf that had otherwise been expected
to have a larger velocity dispersion \citep{collins13}.

\subsubsection{And XXII}

And XXII was observed by \citet{tollerud2012}, albeit with relatively few (7) stars.  \citet{collins13} observe ten stars.
Both independent observations agree with the MOND predictions.
Note that this case spans the transition between EFE and isolated regimes \citep{milg7dw},
and has a small enough velocity dispersion that the contribution of binaries may be relevant.

\subsubsection{And XXIII}

And XXIII has a velocity dispersion based on 42 stars \citep{collins13}, the most of any dwarf of Andromeda to date.
Unfortunately, the photometric data are somehwat uncertain, limiting the accuracy of our prediction.
The MOND prediction is consistent with the observed velocity dispersion, but improved photometric data would
sharpen the test in this case.  If nothing changes as the error bars shrink, this object would require a mass-to-light
ratio at the upper end of the range we consider.

\subsubsection{And XXIV}

\citet{collins13} provide an upper limit of $\s_{obs} \le 7.3\;\kms$ for And XXIV.  This is safely above our prediction of only a few \kms.
We continue to anticipate a value in this vicinity as the kinematic data improve.

\subsubsection{And XXV}

And XXV is a well observed case based on 26 stars \citep{collins13}.
This observation is in good agreement with the prediction of \mmmm.
And XXV is one of the cases that \citet{collins13} note as being an outlier,
but MOND correctly anticipates its relatively low velocity dispersion.

\subsubsection{And XXVI}

The predicted velocity dispersion for And XXVI is lower than the observed value.
The latter is based on only 6 stars \citep{collins13}, so we do not yet consider this case to be problematic.
It would become so if the errors shrank without a decline in $\s$ as happened with And XXI.

\subsubsection{And XXVII}

And XXVII is an interesting case.
It is predicted to have a velocity dispersion of only $\approx 2\;\kms$.
\citet{collins13} observe a much larger value ($\approx 15\;\kms$).
However, they point out that this system is not in equilibrium, rendering it unfit for analysis in the present context.

And XXVII appears coincident with one of the streams around M31.
This is intriguing since tidal forces in MOND should be relatively more effective at perturbing small dwarfs \citep{bradadwarf}
than Newtonian gravity when the dwarf is cocooned in a dark matter halo \citep{pendwA}.
We therefore anticipate some associations of dwarfs with streams \citep{mw10}.

\subsubsection{And XXVIII}

\mmmm\ highlighted And XXVIII as perhaps the best test case among those dwarfs that did not yet have velocity dispersion measurements.
At a large distance from M31, it is most clearly in the isolated regime.
This is helpful because it reduces the potential for systematic errors.  The predicted velocity dispersion depends
only on the stellar mass in the isolated regime, and not on the half-light radius or the external field of M31.
Moreover, the $\s \sim M^{1/4}$ dependence in the isolated regime is less sensitive to the assumed range of mass-to-light ratios.

Our predicted velocity dispersions were 3.6, 4.3, and $5.1\;\kms$ for $\ML =1$, 2, and $4\;\MLsun$, respectively.
Subsequent observations find $4.9\pm1.6\;\kms$ \citep[based on 18 stars]{tollerud2013}
and $6.6^{+2.9}_{-2.1}\;\kms$ \citep[based on 17 stars]{collins13}.
Our narrow range of predicted velocity dispersion is consistent with both independent sets of data.

\begin{deluxetable*}{lcccccc}
\tablewidth{0pt}
\tablecaption{Predicted Velocity Dispersions}
\tablehead{
\colhead{Dwarf} & \colhead{$L_V$} & \colhead{$r_{1/2}$} & \colhead{$\sigma_{iso}$} & \colhead{$\sigma_{efe}$} & \colhead{$g_{in}$} \\
& \colhead{$10^5\;L_{\sun}$} & pc & \multicolumn{2}{c}{\kms} &\colhead{$a_0$}
}
\startdata
And XXX   & \phn1.4\phn & \phn356 & $3.8^{+0.7}_{-0.6}\pm0.7$  & $3.5^{+1.5}_{-1.0}\pm1.3$ & 0.013 \\
And XXXI   & 41.\phn             & 1216 & $9.0^{+1.7}_{-1.4}\pm1.5$  & \dots & 0.024  \\ 
And XXXII   & 71.\phn             & 1941 & $10.3^{+1.9}_{-1.6}\pm1.7$ & $10.3^{+4.3}_{-3.0}\pm3.5$ & 0.020
\enddata
\tablecomments{Predictions for And XXX/Cass II are based on the data of \citet{collins13} and those for
And XXXI/Lac I and And XXXII/Cass III are based on the data of \citet{martin13}.
The first uncertainty is that from a factor of two in the mass-to-light ratio and the second is that from observational errors,
as per Table~\ref{origpredtable}.  The characteristic acceleration at the half-light radius is given in the last column.}
\label{morepredtable}
\end{deluxetable*}

\subsubsection{And XXIX}

And XXIX spans the EFE and isolated regimes, so its prediction is less secure than that of And XXVIII.
Our prediction of $\s = 4.9\;\kms$ for $\ML=4\;\MLsun$ is within the observed range of
$5.7\pm1.2$ \citep[based on 24 stars]{tollerud2013}.  The lower assumed $\ML$ values predict velocity dispersions below the observed range,
though $\ML = 2\;\MLsun$ is acceptable given the photometric uncertainties.

\subsection{Systematics Effects}

Many potential systematic uncertainties were discussed in \mmmm, and we will not revisit that discussion here \citep[see also][]{mw10}.
We do note that the predicted velocity dispersions of many dwarfs are consistent with but slightly lower than the observed values.
This is not obviously significant given the uncertainties.  Nevertheless, it is striking that of all the dwarfs
considered here, only And XV has a predicted velocity dispersion that is slightly higher than the observed value.
One might infer from this that the mean mass-to-light ratio is on the high side of our assumed range.
This might indeed be the case.  But we also expect that most systematic effects will cause the observed velocity dispersion to be
overestimated.  This can be caused by interlopers or binary stars, so rather than requiring a high mass-to-light ratio,
we might be seeing a hint of such an effect.  If binary stars contribute even a few \kms\ in quadrature to the observed
velocity dispersion, it would noticeably lower the inferred \ML.
A comprehensive treatment of the effects of binary stars and interlopers is beyond the scope of this work \citep[see][]{angusdw,serradw}.

\subsection{Comparison of Pairs of Similar Dwarfs in the Isolated and EFE Regimes}
\label{paircomp}

The distinction in MOND between the EFE and isolated regimes provides us with extra leverage to test the theory.
In the isolated regime, only the luminous mass matters ($\s \propto M^{1/4}$).
In cases where the EFE dominates, the distribution of that mass (as encoded by $r_{1/2}$) also matters [$\s \propto (M/r_{1/2} g_{ex})^{1/2}$].
Consequently, two dwarfs with the same mass will have a different predicted velocity dispersion in these different regimes.

And XVII and And XXVIII are examples of two dwarfs with similar luminosities ($\sim 2 \times 10^5\;\Lsun$) and scale sizes ($\sim 300$ pc).
If both were in the same regime, they should have indistinguishable velocity dispersions.
Despite their similarity, And XVII is in the EFE regime while And XXVIII is isolated because of their different distances from M31.
Consequently, And XVII is predicted to have a smaller velocity dispersion ($\approx 2.5$ instead of $\approx 4.5\;\kms$).
The difference between the two regimes is nearly a factor of two, but also only $\sim$$2\;\kms$ --- a rather subtle distinction to make.
Nevertheless, such a difference is shown by the new observations ($\approx 3\;\kms$ for And XVII and $\approx 5\;\kms$ for And XXVIII).

Another example is provided by And XVIII and And XXV.  Both have similar luminosities ($\approx 6\times10^5\;\Lsun$) but scale
radii that differ by a factor of $\sim 2$.  The larger size of And XXV helps place it in the EFE regime where its velocity dispersion is
predicted to be $\approx 3.5\;\kms$ instead of $\approx 5.7\;\kms$.  While the prediction for And XXV is in good agreement with recent data,
the situation for And XVIII is less obvious.  If we adopt the observation with the larger number of stars, then the measured velocity dispersion
is indeed larger as expected.

There are more pairs of similar dwarfs in \mmmm, though most of these pairs happen to have both members in the same regime.  
Perhaps the most interesting pair is And XVI and And XXI.  Both have similar luminosity ($\approx 4 \times 10^5\;\Lsun$) but quite
different size ($r_{1/2} = 178$ pc for And XVI but 1023 pc for And XXI).  Consequently, And XVI is isolated while And XXI is in the EFE regime.
The resulting distinction in velocity dispersion is modest, and not yet distinguished by the data.

The distinction between isolated and EFE regimes has the potential to provide an important test.
The distribution of stars matters to the characteristic velocity dispersion in the EFE case while it does not in the isolated case.
This not only helps to test MOND, but also to distinguish it from \LCDM\ where one would not anticipate
the same dependence on the distribution of stellar mass.

\section{Further Predictions}
\label{newdwarfs}

This field is growing at a remarkable rate.  Not only have velocity dispersion measurements appeared for the dwarfs discussed above within weeks of the
predictions of \mmmm, but new dwarfs continue to be discovered.  In this section we predict the velocity dispersions of three recently identified dwarfs: 
And XXX \citep[also known as Cass  II]{collins13}, And XXXI, and And XXXII
\citep[also known as Lac I and Cass III, respectively]{martin13}.
We follow exactly the same procedure as described in \mmmm.
Predictions for these objects are reported in Table~\ref{morepredtable}.

\citet{collins13} provide an initial velocity dispersion estimate of $11.8^{+7.7}_{-4.7}\;\kms$ for And XXX.
This is rather larger than our predicted value of $\approx 4\;\kms$.
However, the significance of this difference is $< 2\s$.
Indeed, the measured velocity dispersion is non-zero at only the $2.5 \s$ level.
The measurement is based on only eight stars, and the analysis is beset by a considerable galactic foreground in both position and velocity space.
This complicates identification of true members, heightening the concern over interlopers contaminating the result.
A further complication in this case may be the proximity of And XXX to NGC 147 and NGC 185.

And XXXI and And XXXII should provide good tests.  Both are relatively bright, quite large ($r_{1/2} > 1$ kpc),
and reasonably remote from Andromeda itself.  The predicted velocity dispersion of both is $\approx 10\;\kms$.
And XXXI is expected to be in the isolated regime, for which our prediction is more precise.
Thanks to its large size ($r_{1/2} \approx 2$ kpc), And XXXII straddles the boundary between isolated and EFE regimes
in spite of its large distance from M31.  Nevertheless, there is not much flexibility in our prediction for either dwarf (Table \ref{morepredtable}).

\section{Discussion and Comparison to \LCDM}
\label{discuss}

By and large, MOND performs quite well in predicting the velocity dispersions of the dwarf spheroidal satellites of Andromeda.
This is achieved with reasonable stellar mass-to-light ratios, eliminating the mass discrepancies in these systems.
In contrast, the Newtonian dynamical mass-to-light ratios exceed ten and frequently approach $\sim 50$.
Such large mass discrepancies were not anticipated by conventional theory.  Only MOND predicted this from the outset: the
low surface brightness of dSph systems imply very low accelerations, necessarily pointing to large mass discrepancies.

This paper has focussed on \textit{a priori} predictions. The other cases discussed in \mmmm\ already had some measurements of the velocity dispersion,
but these in no way inform the  MOND predictions. In only a few cases are the data somewhat in tension with the predictions.
Given the nature of astronomical data and the potential for systematic errors, it would be surprising if there were not a few such cases.

A pertinent question is how this compares to the dark matter paradigm.
What does \LCDM\ predict?
Why is MOND-like phenomenology \citep{M04} ever observed in \LCDM?
Is the predictive success of MOND a problem for the dark matter picture \citep[e.g.,][]{sandersfalsify}?

The \LCDM\ paradigm does not make clear and inescapable predictions as MOND does for individual dSph galaxies.
The velocity dispersions is determined almost wholly by the dark matter, so it is not obvious that the luminosity
should have much predictive power.  For example, the work of \citet{Guo10} implies $\s \propto L^{0.1}$  

Most baryons in these tiny dwarfs must either be missing (i.e., blown out by feedback) or themselves be dark.
This is not a subtle effect: $> 99\%$ of the primordially available baryons fail to form stars \citep{M10}.
A 1\% variation in the number of baryons expelled is the difference between a dwarf that is twice as bright as typical
and one that has no stars at all.  It is unlikely that the scatter resulting from any expulsion mechanism can be so small \citep[e.g.,][]{BNA13}, 
so it is remarkable that a procedure based on the luminous mass like the one we have employed should provide detailed, accurate predictions

There have of course been efforts to predict the properties f dwarf satellites in the context of \LCDM.
\citet{PMN08} find that halo mass is not simpoly related to luminosity, as we just argued ourselves.
They also predict that larger dwarfs will have larger velocity dispersions at a given luminosity.
This follows essentially from the stars probing a larger portion of the rising rotation curve of the dark matter halo.
As discussed in \S \ref{indcase}, there are now examples of large dwarfs with small velocity dispersions, in
contradiction to this expectation.  This is an important instance in which the predictions of
\LCDM\ and MOND are clearly distinct.  It is the prediction of MOND that is realized in the data.

\section{Conclusions}
\label{conclude}

We have used new data to test our published predictions (\mmmm) for the velocity dispersions of the dSph satellites of Andromeda.
The \textit{a priori} predictions fare well.  MOND is not merely consistent with the data, it has the ability to predict it in advance.
This is particularly notable in the case of several large, diffuse dwarfs for which MOND correctly anticipated small velocity dispersions
of only a few \kms\ when large dispersions ($\gtrsim 10\;\kms$) had been anticipated.

We note that a further test is provided by the dichotomy between the isolated and external field dominated cases in MOND.
In the isolated regime, the predicted velocity dispersion depends only on the luminous mass. 
In the regime where the external field of the host dominates, the velocity dispersion also depends on the size of the dwarf and the strength of
the external field.  Consequently, photometrically identical dwarfs are predicted to have different velocity dispersions if one is in the isolated
regime and the other is influenced by the external field imposed by the host galaxy.  
A comparison of similar pairs of dwarfs shows tentative evidence for this dichotomy, which is not expected in \LCDM.




\end{document}